\documentclass[twocolumn,preprintnumbers,amsmath,amssymb,prb]{revtex4}
\usepackage{dcolumn}
\usepackage{bm}
\usepackage[dvips]{graphicx}
\usepackage{subfigure}
\usepackage{longtable}
\usepackage[colorlinks=true, linkcolor=blue, citecolor=blue]{hyperref}
\usepackage{dcolumn}

\begin{document}

\title{First-principles study of lithium-doped carbon clathrates under pressure}

\author{N. Rey$^a$}
\author{A. Mu\~noz$^b$}
\author{P. Rodr\'iguez-Hern\'andez$^b$} 
\author{A. San~Miguel$^a$}

\affiliation{$^a$Laboratoire de Physique de la Mati\`ere Condens\'ee et Nanostructures, Universit\'e Lyon 1; CNRS, UMR 5586, F-69622 Villeurbanne, France}
\affiliation{$^b$Departamnto de Fisica Fundamental II, Universidad de La Laguna
E38205 La Laguna, Tenerife SPAIN  and 
MALTA Consolider Ingenio 2010 team}





\date{\today}

\begin{abstract}
We present a theoretical study on the behavior under pressure of the two hypothetical C$_{46}$ and Li$_8$C$_{46}$ type-I carbon clathrates in order to bring new informations concerning their synthesis. Using \textit{ab initio} calculations, we have explored the energetic and structural properties under pressure of these two carbon based cage-like materials. These low-density meta-stable phases show large negative pressure transitions compared to diamond which represent a serious obstacle for their synthesis. However, we evidence that a minimum energy barrier can be reached close to 40~GPa, suggesting that the synthesis of the Li-clathrate under extreme conditions of pressure and temperature may be possible. Electronic band structure with related density of states behavior under pressure as well as the dependence of the active Raman modes with pressure are also examined.

\end{abstract}


\pacs{Valid PACS appear here}

\maketitle

\section{Introduction}
Hypothetical carbon analogues to silicon clathrates are predicted to exhibit promising properties of fundamental and technological interest. These low-density class of cage-like materials, related to the zeolite topology, were first investigated by Nesper \textit{et al}~\cite{nesper93}. They allow the intercalation of guest atoms inside the cages and therefore display unique properties. Exceptional values of mechanical properties or superconductivity at high temperature are predicted within the framework of \textit{ab initio} calculations~\cite{nesper93,saito97,blase04,connet03}. Also the low work function of the metallic Li$_8$C$_{46}$ clathrate would  be interesting in the design of potential electron-emitting devices~\cite{timo02}. Contrarely to the other group-14 clathrates (Si, Ge and Sn), the synthesis of the carbon clathrates remains a challenge up to now even if the deposition of C$_{20}$ or C$_{36}$ networks, which basic units have a comparable size to the ones of the carbon clathrates, have been prepared~\cite{iqbal03,piskoti98}. 

\begin{figure}[!hbt]
\begin{center}
\includegraphics[scale=0.7]{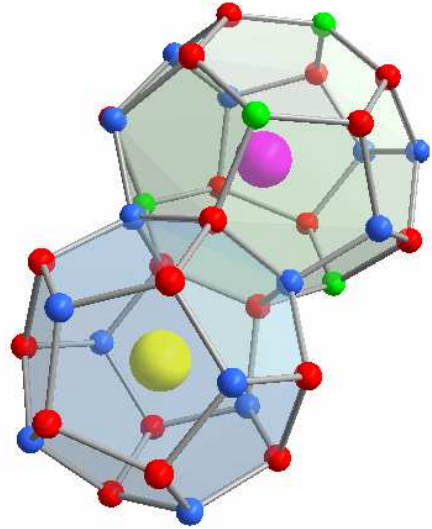}
\end{center}
\caption{\label{fig:lic}Symbolic representation of the C$_{20}$ and C$_{24}$ face-sharing carbon nano-cages.}
\end{figure}

Type-I semiconducting carbon clathrates consist of the assembly face-sharing C$_{20}$ and C$_{24}$ nano-cages resulting in a cubic structure (SG: $Pm\overline{3}n$, No 223) (Fig.~\ref{fig:lic}). They would represent a low-density fourfold diamond-like $sp^3$ configuration. Intercalation with the Li and Be small atoms inside each cage of the type-I carbon clathrate has been suggested~\cite{ker02}. \textit{Ab initio} calculations indicate that upon efficient doping, carbon clathrates would turn metallic~\cite{saito95,saito97,berna00,timo02,spagno03}. Furthermore, large values of the electron-phonon interaction potential let envisage carbon clathrates as possible candidates for high-temperature superconductivity~\cite{saito95,connet03,spagno03}. Recently, Zipoli \textit{et al} have calculated a large value of the electron-phonon coupling for the clathrate FC$_{34}$ leading to a superconducting temperature close to 77~K~\cite{zipoli06}. In addition, X. Blase \textit{et al} have shown that the minimum strength of the carbon clathrates should be larger than that of diamond~\cite{blase04}. C-clathrates have been also predicted to have extraordinary mechanical properties as a bulk modulus close to the one of diamond~\cite{saito97,sanmig99} or even high strength limits under tension and shear beyond diamond ones~\cite{blase04}

The silicon type-I clathrate Ba$_8$Si$_{46}$ can be synthesised from a precursor consisting of a stoichiometric mixture of BaSi$_2$ and Si at 3 GPa and 800$^\circ$C~\cite{yama00}. An attempt was made by Yamanaka \textit{et al} to prepare a carbon clathrate using high pressure (HP) and high temperature (HT) conditions from the polymerization of C$_{60}$ but no evidence of a clathrate-like phase analogous to Si-clathrates were found~\cite{yaman06}. Another possible HPHT route has been envisaged using a Li-graphite intercalation compound, where the Li atoms could frustrate the graphite-diamond conversion and promote the formation of the endohedrally doped nano-cages~\cite{sanmig99,STou05}. This idea is supported by the observation of the half conversion of the $sp^2$ bonds into $sp^3$ bonds at 17~GPa when graphite is pressurized~\cite{mao03}.      

In this work, we investigate the high-pressure properties of the C$_{46}$ and Li$_8$C$_{46}$ carbon clathrates within the framework of the Density Functional Theory as a way to obtain new informations concerning their synthesis. The diamond-clathrates transition is discussed as well as the prospect of a  reaction path leading to the Li-intercalated carbon clathrates. The electronic band structure properties of C$_{46}$ and Li$_8$C$_{46}$ under pressure are also examined. We have calculated the active Raman modes of Li$_8$C$_{46}$ and studied their pressure dependence which could contribute to the identification of carbon clathrate crystals if synthesized for instance in a diamond anvil cell.

\section{Computational details}
Our \textit{ab initio} calculations were performed with the VASP code~\cite{vasp}, using the Projector Augmented Waves (PAW)~\cite{blochl94} pseudopotentials supplied with the package~\cite{kresse99}, within the Generalized Gradient Approximation (GGA) for the exchange-correlation energy~\cite{perdew96}. This choice is justified as in the case of $sp^3$ carbon based materials, GGA gives a better evaluation of the C-C distance~\cite{lee97,janotti01}. A plane-wave energy cutoff of 650 eV  together with a (4$\times$4$\times$4) Monkhorst-Pack grid corresponding to 10 k-points in the irreducible part of the Brillouin zone were used for the C$_{46}$ and Li$_8$C$_{46}$ compounds. Our k-mesh was enough to reach the convergence due to the large clathrate cells. Thus, convergence of 2 meV/atoms in the difference of total energies and 0.2~GPa in pressure was reached. The cell and internal parameters were fully relaxed in order to obtain forces over the atoms lower than 0.005~eV/\AA. The resulting energies as a function of the volume were fitted using a Murnaghan equation of state (EOS)~\cite{murnag44} to obtain the equilibrium volume, the bulk modulus and its pressure derivative at zero pressure. The pressure calculated values are in excellent agreement with the ones provided from the fit. The study of the enthalpy allowed us to determine the relative stable phase.

\section{Results}
\subsection{Structural properties}
\begin{figure}
\includegraphics[trim = 0mm 0mm 0mm 0mm, clip,width=0.45\textwidth]{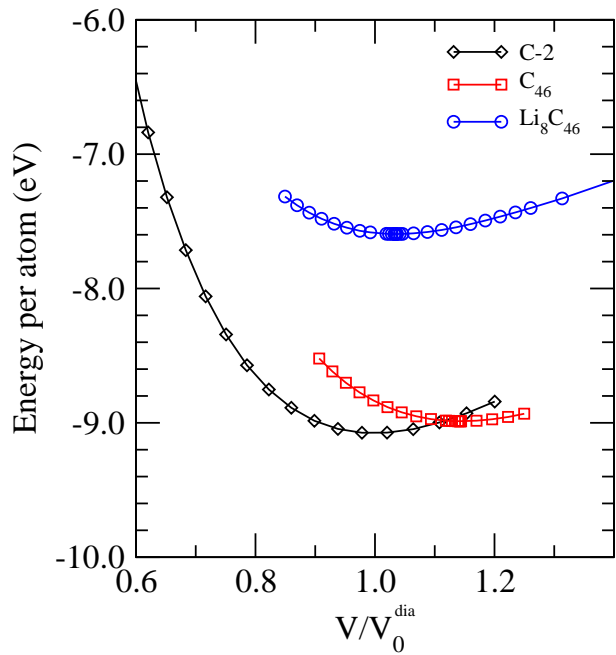}
\caption{\label{fig:energy}Calculated Energy-volume curves for diamond (C$_2$), C$_{46}$ and Li$_8$C$_{46}$. $V_0^{dia}$ is the atomic volume of the diamond structure at equilibrium.} 
\end{figure}
Figure~\ref{fig:energy} shows the calculated energy-volume (E-V) curves of the carbon phases considered, i.e.\ diamond (C$_2$) which is included for comparison and the C$_{46}$ and Li$_8$C$_{46}$ clathrates structures. The diamond phase has the lowest energy and is the most stable at zero pressure. In Table~\ref{tab:res}, the results for the three studied structures are given from our calculations together with existing experimental values. The internal parameters of the equilibrium volume of C$_{46}$ and Li$_8$C$_{46}$ are shown in Table~\ref{tab:intern}. Our calculated values for the lattice parameter and the bulk modulus of diamond are in excellent agreement with the experimental works~\cite{mcskimin72,lb91}. The resulting equilibrium lattice parameter and bulk modulus for C$_{46}$ (a=6.696 \AA, $B_0$=371 GPa) are close to those from previous LDA and GGA studies~\cite{timo02,zipoli06}. The intercalation of Li atoms at the center of the C$_{20}$ and C$_{24}$ cages leads to an expansion of the lattice parameter (a=6.833 \AA) as well as a higher compressibility ($B_0$=356 GPa). Atomic volume has been slightly expanded by 14.5 \% and 3.5 \% for the C$_{46}$ and the Li$_8$C$_{46}$ phases respectively with respect to diamond. We find an energy difference of 92 meV/atom for C$_{46}$ relative to diamond in good agreement with Nesper \textit{et al} (90 meV/atom)~\cite{nesper93} who used also the PAW method. Other \textit{ab initio} methods give much higher energy differences 144-210 meV/atom~\cite{adams94,benedek95,okeeffe98,perottoni01}. Despite the discrepancies found from all the mentioned calculations, it is remarkable that these binding energies, relative to diamond, are considerably lower than that of the fullerene C$_{60}$ (430~meV/atom)~\cite{adams92}. However, according to our calculations, filling the voids inside the C$_{46}$ clathrate dramatically increases the enthalpy at zero pressure to 1.48 eV/atom.  
\begin{table}
\caption{\label{tab:res}Structural and cohesive parameters of diamond (C$_2$), C$_{46}$ and Li$_8$C$_{46}$. The equilibrium volume $V_0$ and cell parameter $a_0$ are obtained after relaxation of the forces. Bulk modulus $B_0$ and its pressure derivative $B'_0$ are obtained with the Murnaghan EOS fit. The relative atomic volume $V/V_0$ and enthalpy $\Delta H_0$ with respect to the diamond phase are also given.}
\begin{center}
\begin{ruledtabular}
\begin{tabular}{ldddldd}
 & \multicolumn{1}{c}{a$_0$} & \multicolumn{1}{c}{$V_0$}  & \multicolumn{1}{c}{$V/V_0$} & \multicolumn{1}{c}{$B_0$} & \multicolumn{1}{c}{$B'_0$} & \multicolumn{1}{c}{$\Delta H_0$} \\
 & \multicolumn{1}{c}{(\AA)} & \multicolumn{1}{c}{(\AA$^3$)} &   &  \multicolumn{1}{c}{(GPa)} & & \multicolumn{1}{c}{(eV/atom)} \\
\hline
C-2& 3.576  & 11.4 & 1.000& 437 & 3.22 & 0 \\
    & 3.567\footnote{Experimental value from reference~\cite{lb91}.}  &   &     &  442\footnote{Experimental value from reference~\cite{mcskimin72}.}      &       &   \\
C$_{46}$ & 6.696 & 300.2 & 1.145 & 371 & 3.44  & 0.09 \\
Li$_8$C$_{46}$ & 6.833 & 319.0 & 1.035 & 356 & 3.13 & 1.48 \\
\end{tabular}
\end{ruledtabular}
\end{center}
\end{table}

\begin{table}
\caption{\label{tab:intern} Internal parameters of the C$_{46}$ and Li$_8$C$_{46}$ clathrates at their equilibrium volume.}
\begin{center}
\begin{tabular}{|c|c|ddd|}
\hline
\multicolumn{5}{|c|}{C$_{46}$, SG: $Pm\overline{3}n$, No 223, Z=1} \\
\hline
Atoms & Wyckoff positions & \multicolumn{3}{c|}{Internal parameters (x, y, z)}\\
C (1) &  $6c$ & 0.25 &  & 0.5 \\
C (2) & $16i$ (x, x, x) & 0.1845 & 0.1845 & 0.1845 \\
C (3) & $24k$ (0, y, z) & 0&  0.3056 & 0.1187 \\
\hline
\multicolumn{5}{|c|}{Li$_8$C$_{46}$, SG: $Pm\overline{3}n$, No 223, Z=1} \\
\hline
C (1) &  $6c$ & 0.25 & 0 & 0.5 \\
C (2) & $16i$ (x, x, x) & 0.1854 & 0.1854 & 0.1854 \\
C (3) & $24k$ (0, y, z) & 0 & 0.0345 & 0.1213 \\
Li (1)& $2a$ & 0 & 0 & 0 \\
Li (2)& $6d$ & 0.25 & 0.5&  0 \\
\hline
\end{tabular}
\end{center}
\end{table}

\begin{figure}
\includegraphics[scale=0.6]{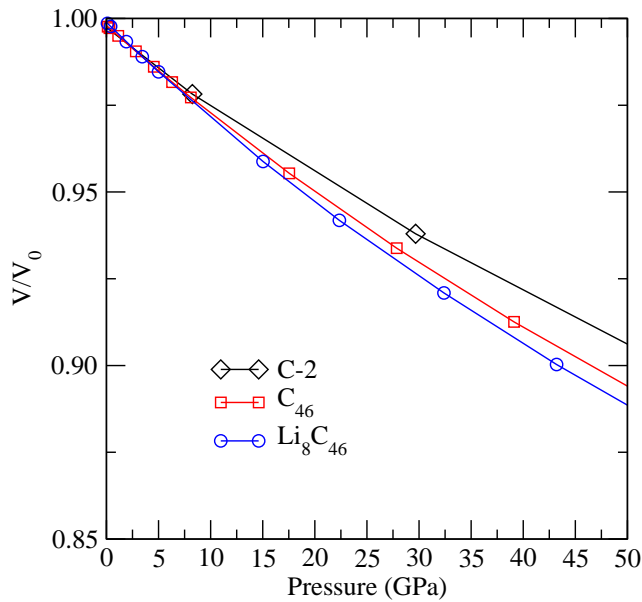}
\caption{\label{fig:vol}Volumic variation as a function of pressure for (C$_2$), C$_{46}$ and Li$_8$C$_{46}$.} 
\end{figure}
\begin{figure}
\includegraphics[trim = 0mm 0mm 0mm 0mm, clip, width=0.45\textwidth]{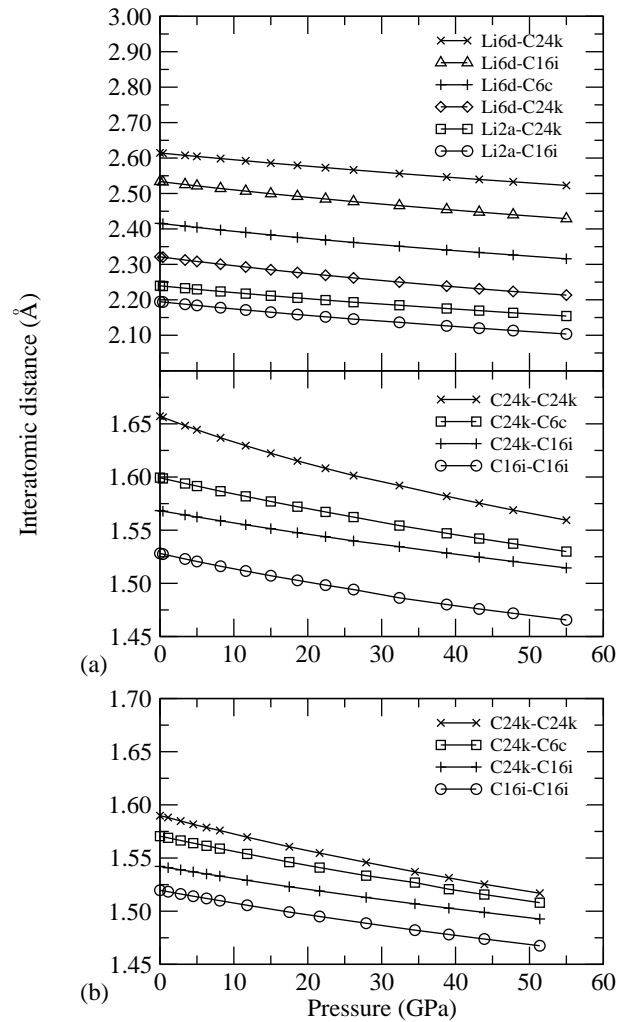}
\caption{\label{fig:inter}Top: Interatomic distances evolution with pressure for Li$_8$C$_{46}$ (a): Li-C (top panel) and C-C distances (bottom panel), and for C$_{46}$ (b): C-C distances.} 
\end{figure}

The compressibility of the C$_{46}$ and Li$_8$C$_{46}$ clathrates is close to the diamond one until $\sim$8 GPa (Fig.~\ref{fig:vol}). Then, C$_{46}$ and Li$_8$C$_{46}$ become slightly more compressible than diamond, the pure clathrate being less compressible than the filled clathrate. This contrasts with the observations in Si-clathrates~\cite{sanmig02} in which intercalation trends to shift the bulk modulus towards the Si-diamond phase. The pressure-evolution of the C-C and Li-C distances in the cages of the two clathrates are displayed in Fig.~\ref{fig:inter}.
The insertion of lithium inside each cages of C$_{46}$ leads to a slight extension of the cages and produces a  general stiffening in the pressure-evolution of the C-C distances. The Li-C and C-C distances decrease smoothly with pressure. There is no sudden drop corresponding to a cell volume reduction like it has been experimentally observed in the Ba-Si and Si-Si distances of Ba$_8$Si$_{46}$~\cite{sanmig05,yang06}.
\subsection{Stability under pressure}
\begin{figure}[!t]
\begin{center}
\includegraphics[width=0.5\textwidth]{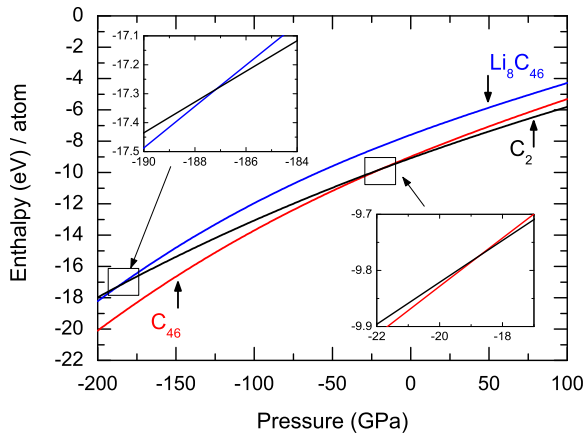}
\end{center}
\caption{\label{fig:enthalpies}Enthalpies of the diamond (C$_2$) and the C$_{46}$ and Li$_8$C$_{46}$ clathrates phases as a function of pressure. The insets provide a zoom of the regions of interest.}
\end{figure}
The diamond structure is the most stable phase studied in this work for positive pressures (Fig.~\ref{fig:enthalpies}). Enthalpy curves of C$_{46}$ and Li$_8$C$_{46}$ are found to be lower than the one of diamond at -19~GPa and -187~GPa, respectively. Our C$_2$$\rightarrow$C$_{46}$ pressure transition agrees reasonably with the one calculated by Perottoni and Da Jornada within an Hartree-Fock approximation~\cite{perottoni01}. Negative pressure values are expected in those kind of transitions implying expanded phases compared to diamond. Although, in absolute values, they are much larger than those calculated in the case of the Si$_2$$\rightarrow$Si$_{46}$ (-6 GPa)~\cite{perottoni01} and Ge$_2$$\rightarrow$Ge$_{46}$ (-2.4 GPa)~\cite{dong99} transitions. In the case of the intercalated type-I carbon clathrates with Na (P$_t$=-77~GPa)~\cite{perottoni01} or Li, huge negative pressure transitions relative to diamond seem to be systematic and render difficult the observation of the doped clathrate phases.
\begin{figure}[!t]
\begin{center}
\includegraphics[width=0.5\textwidth]{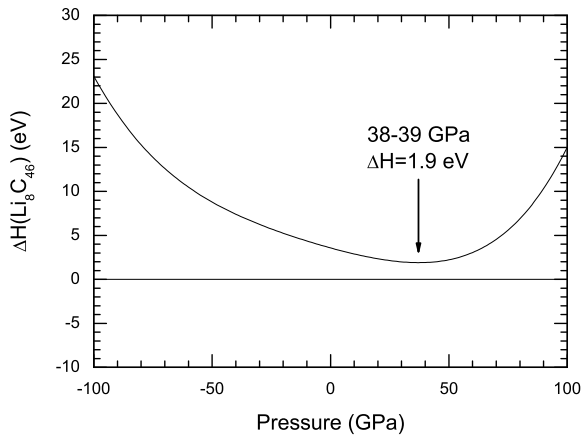}
\end{center}
\caption{\label{fig:deltaH}Relative enthalpy of Li$_8$C$_{46}$ from Eq.~\ref{eq:deltaH} as a function of pressure. $\Delta H(Li_8C_{46})>$0 for the pressure range studied. A minimum corresponding to an energy barrier of 1.9~eV is found at 38-39~GPa.}
\end{figure}

In order to hedge the phase stability of Li$_8$C$_{46}$, we considered the following reaction:
\begin{equation}
8\:Li+23\:C_2\rightarrow Li_8C_{46},
\label{eq:reac}
\end{equation}
involving stoichiometric ratio of metallic lithium (Li) and diamond (C$_2$) to obtain the Li$_8$C$_{46}$ compound. Thus, we studied the variation of enthalpy of the previous reaction, defined as follows
\begin{equation}
\Delta H(Li_8C_{46})=H(Li_8C_{46})-8H(Li)-23H(C_2). 
\label{eq:deltaH}
\end{equation}

Figure~\ref{fig:deltaH} shows the variation of $\Delta H(Li_8C_{46})$ as a function of the pressure. $\Delta H(Li_8C_{46})$ is always positive indicating that the considered reaction~\ref{eq:reac} is not favourable in the pressure range studied. However, we have found a minimum corresponding to an energy barrier of 1.9~eV around 38-39 GPa.

\subsection{Electronic properties under pressure}
\begin{figure*}[!tb]
\begin{center}
\includegraphics[trim = 0mm 0mm 0mm 0mm, clip, width=0.8\textwidth]{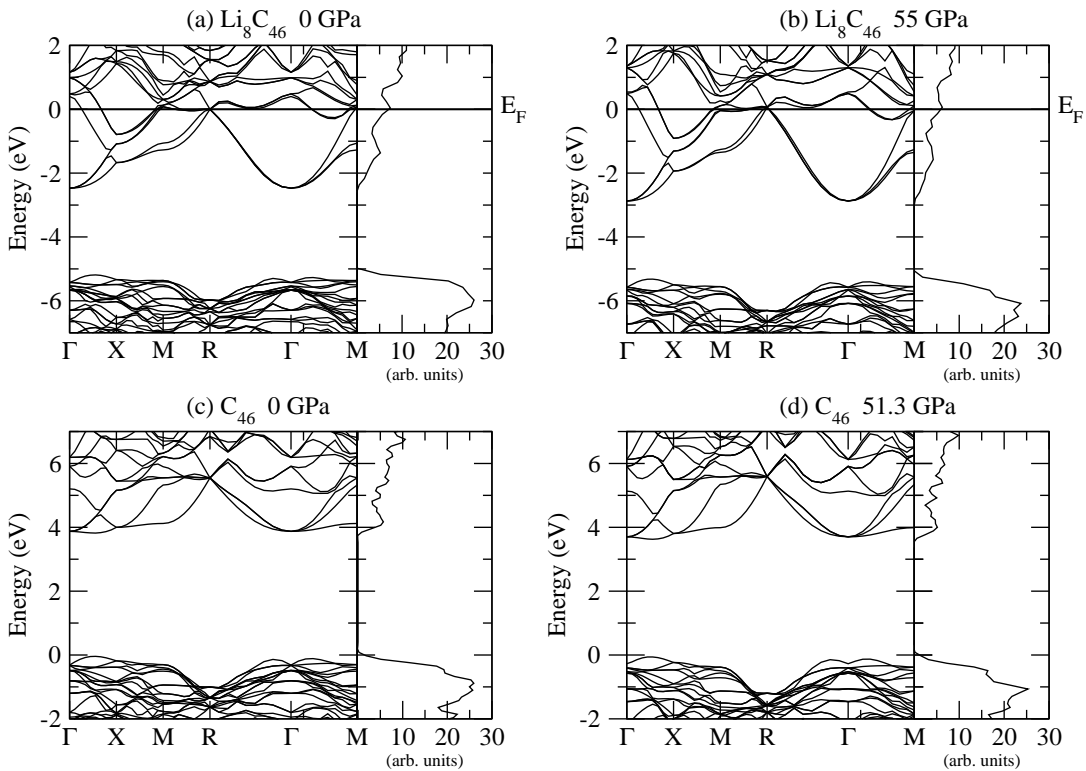}
\end{center}
\caption{\label{fig:bdos}Band structure and density of states associated for Li$_8$C$_{46}$ at 0 GPa (a) and at 55 GPa (b) and for C$_{46}$ at 0 GPa (c) and at 51.3 GPa (d).In the case of the Li$_8$C$_{46}$ metallic compound, the Fermi energy is taken as reference. In the semiconducting C$_{46}$, the top of the valence band is chosen as the energy reference.}
\end{figure*}
We have calculated the electronic band structure of pure and doped carbon clathrates choosing the same path along the same high-symmetry directions of the Brillouin zone as in Ref.~\onlinecite{connet04} were chosen. The band structure of the semiconducting C$_{46}$ at zero pressure and at 51.3~GPa is provided in Fig.~\ref{fig:bdos}(c) and (d). We find a quasi-direct band gap of approximatively 3.88~eV at zero pressure between $\Gamma$-X which is in good agreement with previously reported results~\cite{timo02,blase03}. The band gap value decreases with pressure, reaching 3.69~eV at 51.3~GPa, and becomes direct from 2.8~GPa. We can check in Figure~\ref{fig:bdos}(a) that the intercalation of lithium in each cage of C$_{46}$ leads to a metallic behavior with the filling of the conduction-band by Li 2$s$ electrons. A maximum in the density of states at the Fermi level $E_F$ is observed suggesting interesting superconducting properties as previously remarked~\cite{saito97,timo02,connet03,spagno03}. No dramatic changes have been found at 55 GPa in the band structure of Li$_8$C$_{46}$, especially in the conduction band at the vicinity of $E_{F}$ (Fig.~\ref{fig:bdos}.(d)). In this regard, a band crossing over the Fermi level in the conduction-band of Ba$_8$Si$_{46}$ near the R-point, ascribed to an electronic topological transition, has been calculated recently~\cite{yang06}. In Li$_8$C$_{46}$, the Fermi energy is found to increase with pressure, corresponding to a shift of $\Delta E_F$ $\approx$ 1.7~eV between 0 and 55 GPa. This indicates that the pressure is a valuable parameter to tune the charge transfer in these guest-host systems. We point out that the density of states around E$_F$ present a maximum at 55 GPa.
\begin{table}
\caption{\label{tab:raman_modes}Raman active modes of Li$_8$C$_{46}$ (point group: O$_h$) for the equilibrium volume (at the $\Gamma$-point). The frequency at 870 cm$^{-1}$ consists in the combination of a Raman (R) and an infra-red (IR) mode. These frequencies and their representation were calculated with the \textsc{phonon} program~\cite{parlinski03}.}
\begin{center}
\begin{tabular}{rcr}
\hline
Representation & Frequency (cm$^{-1}$) & Activity \\
\hline
T$_{2g}$  &  247        & R \\
E$_g$     &  310        & R \\
E$_g$     &  387        & R \\
T$_{2g}$  &  425        & R \\
E$_g$     &  550        & R \\
T$_{2g}$  &  649        & R \\
E$_g$  &     654     & R \\
A$_{1g}$ &   674     & R \\
E$_g$  &    690      & R \\
T$_{2g}$  &  766        & R \\
E$_g$  &   804       & R \\
T$_{2g}$  &  813        & R \\
T$_{2g}$  &   864       & R \\
A$_{1g}$+T$_{1u}$ &  870        & R + IR \\
T$_{2g}$  &   921      & R \\
E$_g$  &    944      & R \\
T$_{2g}$  &   959       & R \\
E$_g$  &   976       & R \\
T$_{2g}$  &  1049        & R \\
A$_{1g}$ &   1080     & R \\
\hline
\end{tabular}
\end{center}
\end{table}
\begin{figure}
\begin{center}
\includegraphics[width=0.5\textwidth]{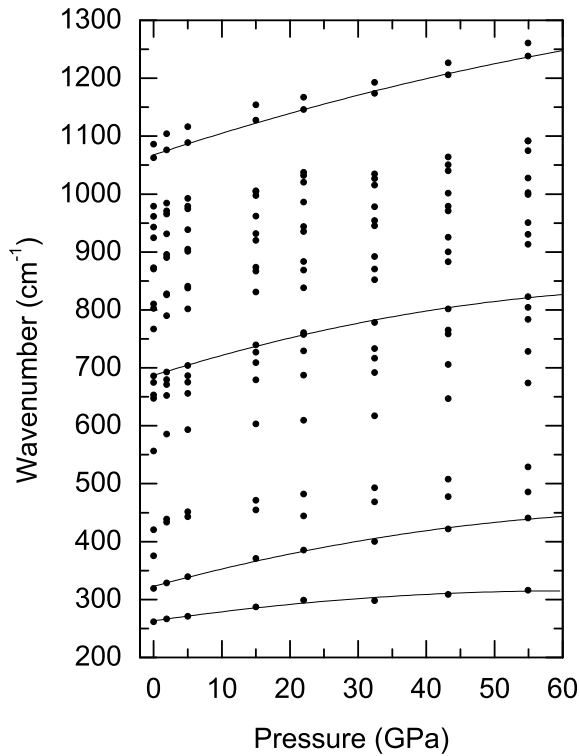}
\end{center}
\caption{\label{fig:modesR}Pressure-evolution of the calculated Li$_8$C$_{46}$ Raman frequencies with the \textsc{phon} program~\cite{alfe98}. The continuous lines represent a polynomial fit (see Table~\ref{tab:fitRam}).} 
\end{figure}
\subsection{Dynamical properties of Li$_8$C$_{46}$ under pressure}
\begin{table*}[!t]
\caption{\label{tab:fitRam} Coefficients of Li$_8$C$_{46}$ Raman frequencies pressure dependence shown in Fig.~\ref{fig:modesR}. A second-order polynomial function was used to fit the calculated frequencies.}
\begin{center}
\begin{tabular}{ldddd}
\multicolumn{5}{c}{$\nu(P)=a+b\;P+c\;P^2$} \\
\hline
 & \multicolumn{1}{c}{T$_{2g}$} & \multicolumn{1}{c}{E$_g$} & \multicolumn{1}{c}{E$_g$}  &  \multicolumn{1}{c}{T$_{2g}$} \\
a (cm$^{-1}$) & 263.28    & 322.83      &686.84  & 1067.68\\
b (cm$^{-1}$/GPa) & 1.69  & 3.18   & 3.68  & 3.87 \\
c (cm$^{-1}$/GPa$^2$) & 1.39\times10^{-2} & 1.95\times10^{-2}& 2.26\times10^{-2} & 1.45\times10^{-2} \\
\hline
\end{tabular}
\end{center}
\end{table*}
We used the direct approach based on the small displacements to calculate the eigenvalues of the dynamical matrix at the $\Gamma$-point of the Brillouin zone~\cite{kresse95,alfe01}. In order to achieve accurate results, we relaxed the cell and the internal parameters to obtain forces over the atoms below 0.0001~eV/\AA. We performed our study with the \textsc{phon}~\cite{alfe98} and \textsc{phonon}~\cite{parlinski03} programs using a (1$\times$1$\times$1) supercell. In order to obtain well converged results, we tested different small displacements, and (4$\times$4$\times$4) k-mesh was used to reach convergence of the phonon frequencies lower than 3-4~cm$^{-1}$. 

In order to link our theoretical results with possible future Raman spectroscopy measurements, if synthesised, we present the list of the active Raman modes of the Li$_8$C$_{46}$ clathrate corresponding to the equilibrium volume in Table~\ref{tab:raman_modes}, that could be helpful as a reference for the experimental study.    

Figure~\ref{fig:modesR} shows the evolution with pressure of the calculated Raman frequencies obtained with the \textsc{phon} program~\cite{alfe98}. We point out that a typical discrepancy of 15~cm$^{-1}$ between the eigenvalues calculated from the two programs \textsc{phon} and \textsc{phonon} was observed. This value can be considered as an average incertitude of our calculated frequencies. The pressure dependence of some frequencies can be reproduced with a second-order polynomial fit as shown with the continuous lines in the figure. For the other frequencies, a more complicated behavior is observed. In that case, we can distinguish three regimes with different effective slopes.  

\section{Summary and conclusions}
In this work, we have presented an \textit{ab initio} theoretical study of the behavior under pressure of the C$_{46}$ and Li$_8$C$_{46}$ carbon clathrates. We have analysed the phase stability of these two compounds with respect to the diamond phase and their structural and electronic properties under pressure. We have also calculated the active Raman frequencies as well as their pressure dependence as a milestone for future experiments. First, our calculations confirmed the interesting structural properties of these clathrates. In particular, these structures are of low compressibility but still more compressible than diamond. As a result, no cell volume collapse with increasing pressure was evidenced like it occurs in silicon clathrate in the same studied pressure range. Our calculations on the C$_{46}$ clathrate are in a good agreement with the previous studies. In the case of Li$_8$C$_{46}$, we find a large enthalpy at zero pressure related to diamond. Furthermore, the Li-intercalated clathrate becomes the stable phase relative diamond near -187 GPa, pointing to a difficulty to synthesize this compound. We have also proposed of reaction path involving stoichiometric ratio of diamond and lithium to obtain the Li$_8$C$_{46}$. Although this reaction seems not be favourable, we have calculated a minimum corresponding to an energy barrier of approximatively 1.9~eV around 38-39~GPa, an easily accessible pressure-window with the diamond anvil cell.
Hence, even if the phase stability of Li$_8$C$_{46}$ appears to be a serious obstacle, our result lets suppose that HPHT conditions are well suited to obtain this compound. \textit{Ab initio} cannot provide a synthesis route, only future attempts will allow to find it. Further \textit{ab initio} calculations on intercalated clathrates considering others chemical species or with different structures will be valuable in order to reduce the cohesive energy.

\begin{acknowledgements}
This work was made possible throught the financial support of the Spanish-French Picasso Program. P. R-H and A.M. acknowledge the financial support of the 
spanish MEC under grants MAT2004-05867-C03-03 and MAT2007-65990-C03-03.
\end{acknowledgements}

\end{document}